\documentclass[
    ,draft            
  ]
  {aipproc}

\layoutstyle{8x11double}

\begin{document}

\title{$^8$Be Nuclear Data Evaluation}

\author{Philip R. Page and Gerald M. Hale\footnote{ E-mail: prp@lanl.gov, ghale@lanl.gov}}{
  address={Theoretical Division, MS B283, Los Alamos National Laboratory,
Los Alamos, NM 87545, USA}
}

\begin{abstract}

An R-matrix analysis of experimental nuclear data on the reactions
$^4He(\alpha,\alpha)$, $^4He(\alpha,p)$, $^4He(\alpha,d)$, $^7Li
(p,\alpha)$, $^7Li (p,p)$, $^7Li (p,n)$, $^7Be (n,p)$, $^6Li
(d,\alpha)$, $^6Li (d,p)$, $^6Li (d,n)$ and $^6Li (d,d)$, leading to
the $^8Be$ intermediate state, has been completed in the last two
years. About 4700 data points from 69 experimental references are
included.  The excitation energy above the $^8Be$ ground state is
$25-26$ MeV for all reactions except $^4He(\alpha,\alpha)$ and $^7Be
(n,p)$. The data for the reactions $^4He(\alpha,\alpha)$ and
$^6Li(d,d)$ do not fit well, but the other reactions fit with a
$\chi^2$/(point) of less than the overall value of $7.9$. Most of the
19 resonances found in the R-matrix analysis correspond to resonances
formerly known from experiment. Evaluated integrated $^4He(\alpha,p)$,
$^4He(\alpha,d)$, $^7Li (p,\alpha)$, $^7Li (p,n)$, $^7Be (n,p)$, $^6Li
(d,\alpha)$, $^6Li (d,p)$ and $^6Li (d,n)$ reaction cross-sections are
presented.  Evaluated cross-section and angular dependence files in
ENDF format were prepared for the twelve reactions $p\:^7Li,\; n\:
^7Be,\; d\: ^6Li \to \alpha\: ^4He,\; p\: ^7Li,\; n\: ^7Be,\; d\:
^6Li$.  Maxwellian averaged temperature-dependent cross-sections in
NDI format were prepared for the six reactions $^7Li(p,\alpha),\;
^7Li(p,n), \; ^7Be(n,p),\; ^6Li(d,\alpha),\; ^6Li(d,p)$ and
$^6Li(d,n)$.

\end{abstract}

\maketitle


This analysis of two-body strong reactions leading to the 
$^8Be$ intermediate state was motivated by large discrepancies
between various evaluations.

There are $16$ different reactions for which cross-sections have been
obtained via this analysis. These are $\alpha\:^4He, \;p\:^7Li,
\;n\:^7Be, \;d\:^6Li \to \alpha\:^4He, \;p\:^7Li, \;n\:^7Be,
\;d\:^6Li$. In addition to unitarity, constraints between reactions
are also provided by time-reversal symmetry (i.e. when the initial and
final particles are interchanged), and isospin symmetry.
For example, no data were entered for
$^4He(\alpha,n)$, $^7Li (p,d)$, $^7Be(n,\alpha)$, $^7Be(n,n)$ and
$^7Be(n,d)$; and only very low energy data for $^7Be(n,p)$.
However,
three of these reactions are strongly constrained via their
time-reverse reactions [$^7Li (p,d)$, $^7Be(n,d)$ and
$^7Be(n,p)$]. Moreover, the other three reactions are 
constrained by isospin symmetry [$^4He(\alpha,n)$ is constrained by
$^4He(\alpha,p)$, $^7Be(n,\alpha)$ by $^7Li(p,\alpha)$, and 
$^7Be(n,n)$ by $^7Li(p,n)$].

Integrated, differential and polarization cross-section 
data were entered for the eleven reactions listed in the abstract.
Substantial data were entered for the $^4He(\alpha,\alpha)$ and 
$^7Li (p,p)$ reactions, and the least data were entered for
the $^4He(\alpha,p)$, $^4He(\alpha,d)$ and $^6Li (d,d)$ reactions.
All reactions where data were entered, except $^4He(\alpha,\alpha)$
and $^7Be (n,p)$, include data up to an excitation energy of $25-26$
MeV. In the $^4He(\alpha,\alpha)$ reaction,
data above the maximum $\alpha$ laboratory energy for which data were
entered ($38.4$ MeV) and below the limit of this analysis ($52$ MeV
laboratory energy), are only available as phase shifts~\cite{bacher},
and have not been incorporated.  For the $^7Be (n,p)$ reaction no data
above the near-threshold data entered were found below the maximum
excitation energy of this analysis ($26$ MeV). 

Cross-sections for the seven reactions mentioned in the abstract in
the energy range corresponding to the excitation energy of this
analysis are shown in Figs. 1-3. The cross-sections are in barns or
millibarns, as indicated; the energies are the laboratory energies
of the projectile. Only the integrated
cross-section data that were entered are indicated. The cross-sections
for $^7Li(p,\alpha)$ and $^6Li (d,\alpha)$ in the Figs. 2-3 should be
divided by 2 to obtain reaction cross-sections, since there are
identical final particles. The shape of the $^4He(\alpha,p)$ (Fig. 1)
reaction is driven by the time-inverse $^7 Li(p,\alpha)$ (Fig. 2)
reaction, which has much more data.  Similarly, the shape of the
$^4He(\alpha,d)$ (Fig. 1) reaction is driven by the time-inverse $^6
Li(d,\alpha)$ (Fig. 3) reaction. Also, the shape of the $^6Li(d,n)$
(Fig. 3) reaction is driven by the $^6Li(d,p)$ (Fig. 3) reaction: the
isospin 0 components of these reactions are related by isospin
symmetry. The magnitude and shape of the $^7Li(p,n)$ cross-section
from $3-7$ MeV have changed considerably at various stages of the
analysis, so that further investigation of this reaction is needed.

\begin{figure*}[p]
 \rotatebox{270}{\scalebox{0.60}{\includegraphics*[30mm,10mm][200mm,250mm]{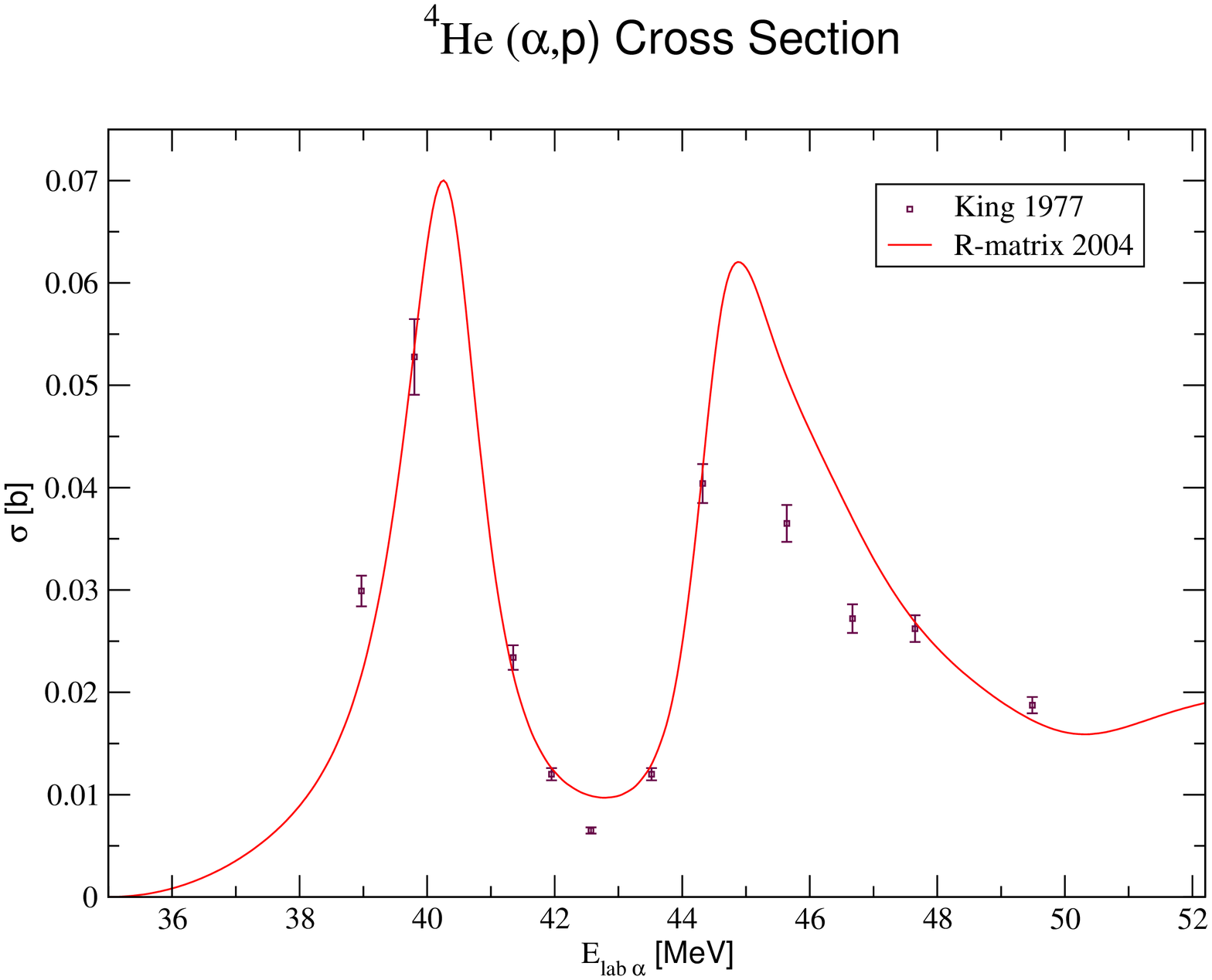}}}
\end{figure*}
\begin{figure*}[p]
 \rotatebox{270}{\scalebox{0.60}{\includegraphics*[30mm,10mm][200mm,250mm]{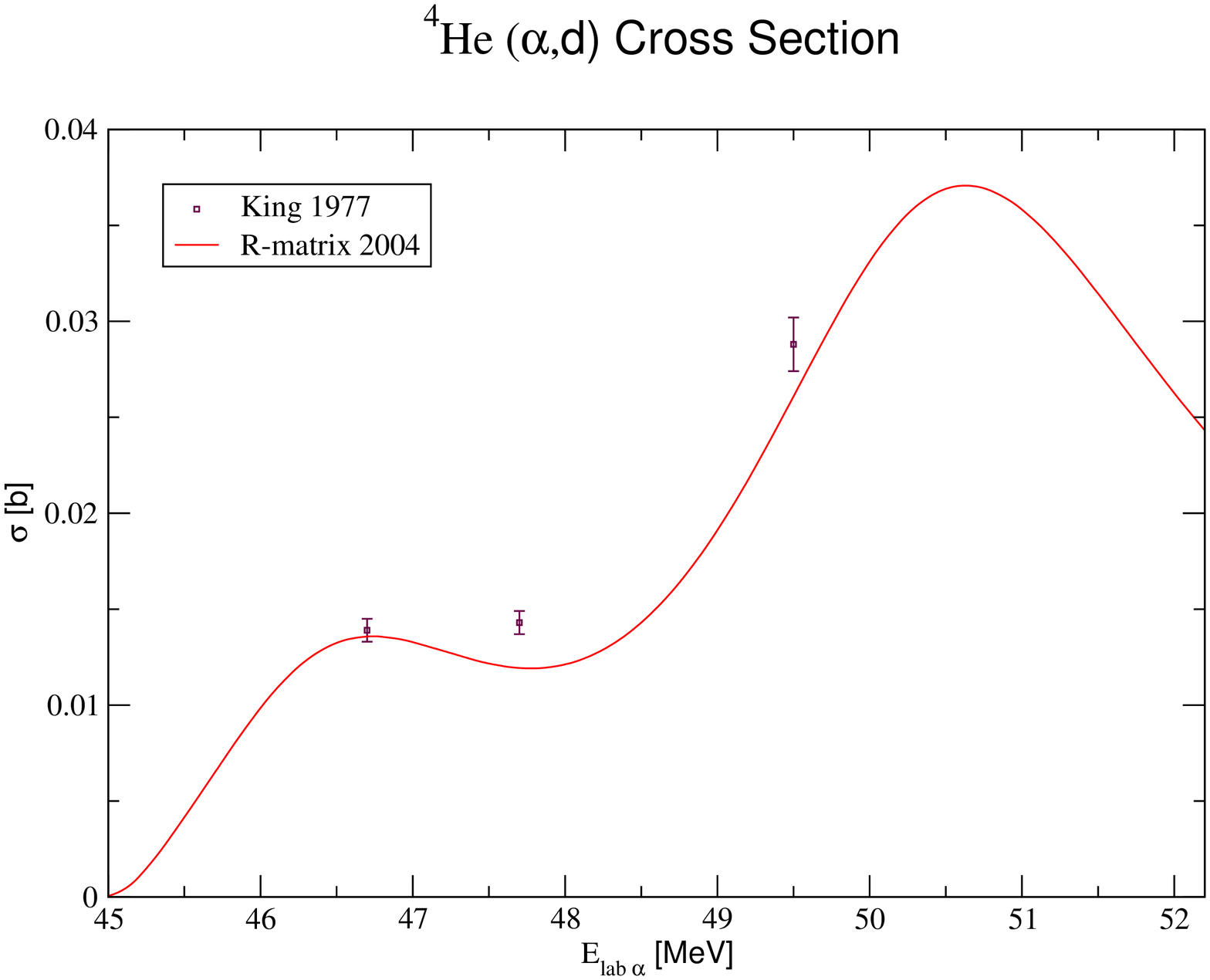}}}
 \caption{From top to bottom: 
Evaluated 2004 R-matrix cross-sections and experimental data for
 (a) $^4He(\alpha,p)$ and (b) $^4He(\alpha,d)$.}
\end{figure*}

\begin{figure*}[p]
 \rotatebox{270}{\scalebox{0.39}{\includegraphics*[30mm,10mm][200mm,252mm]{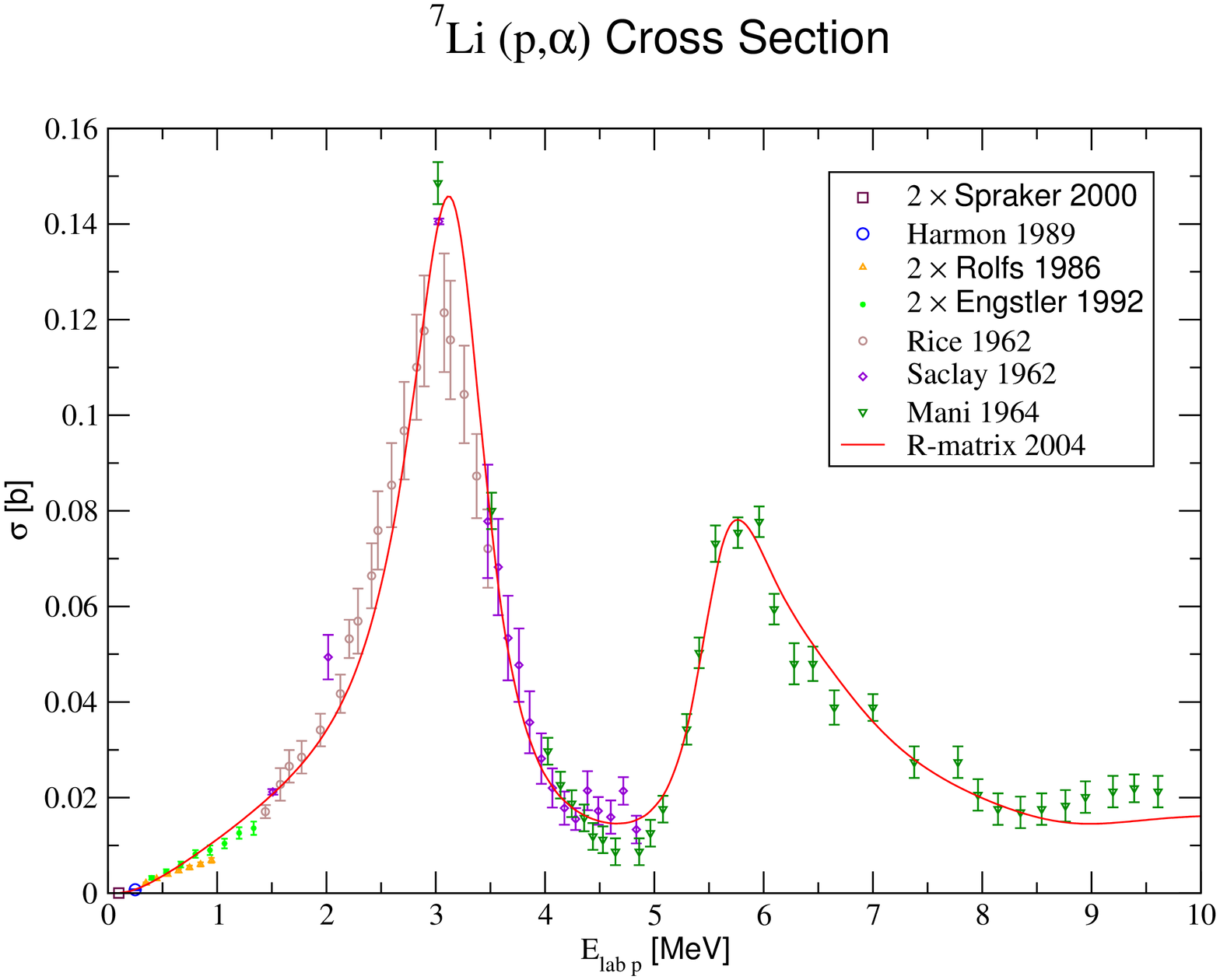}}}
\end{figure*}
\begin{figure*}[p]
 \rotatebox{270}{\scalebox{0.39}{\includegraphics*[30mm,10mm][200mm,250mm]{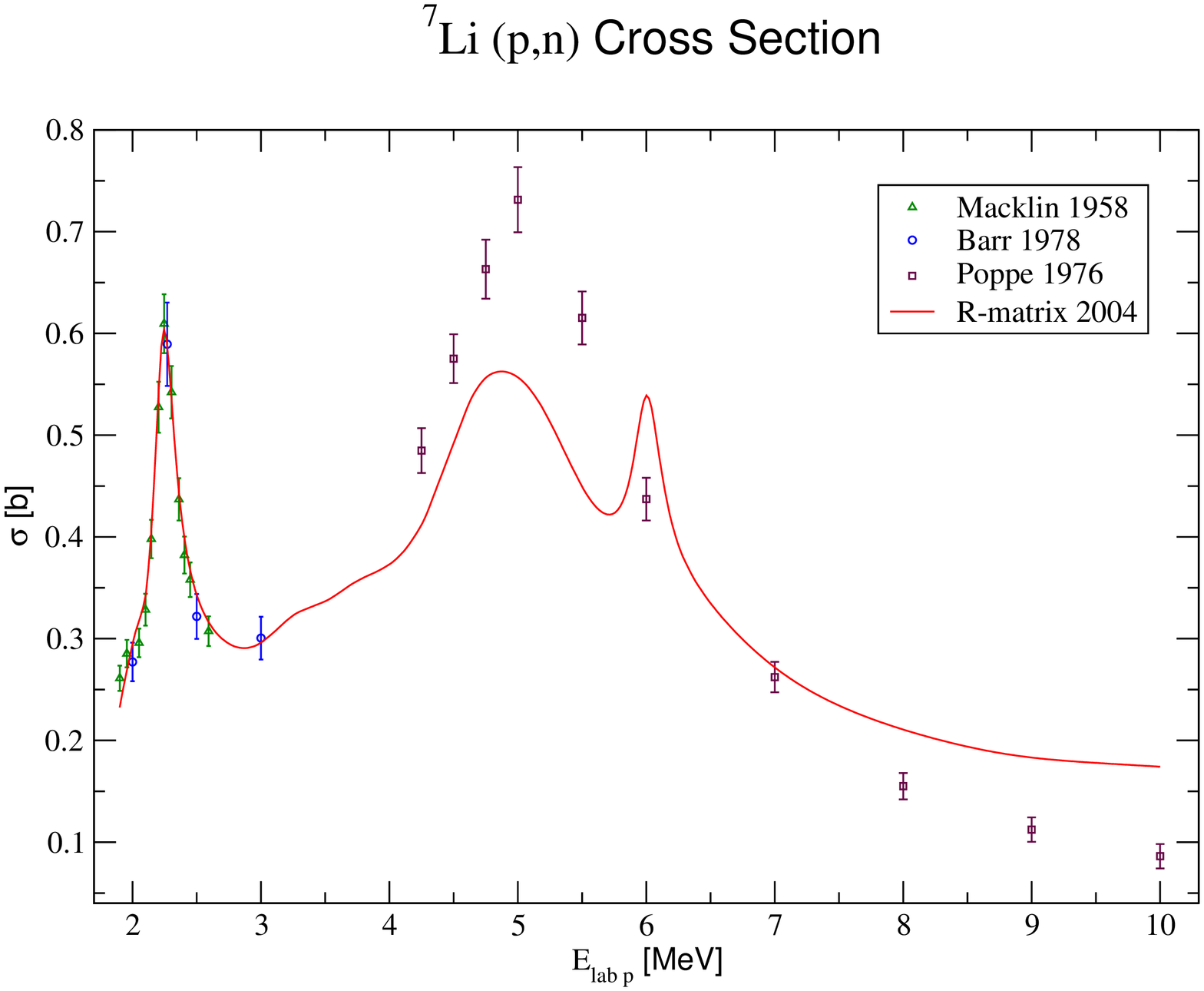}}}
\end{figure*}
\begin{figure*}[p]
 \rotatebox{270}{\scalebox{0.39}{\includegraphics*[30mm,7mm][200mm,250mm]{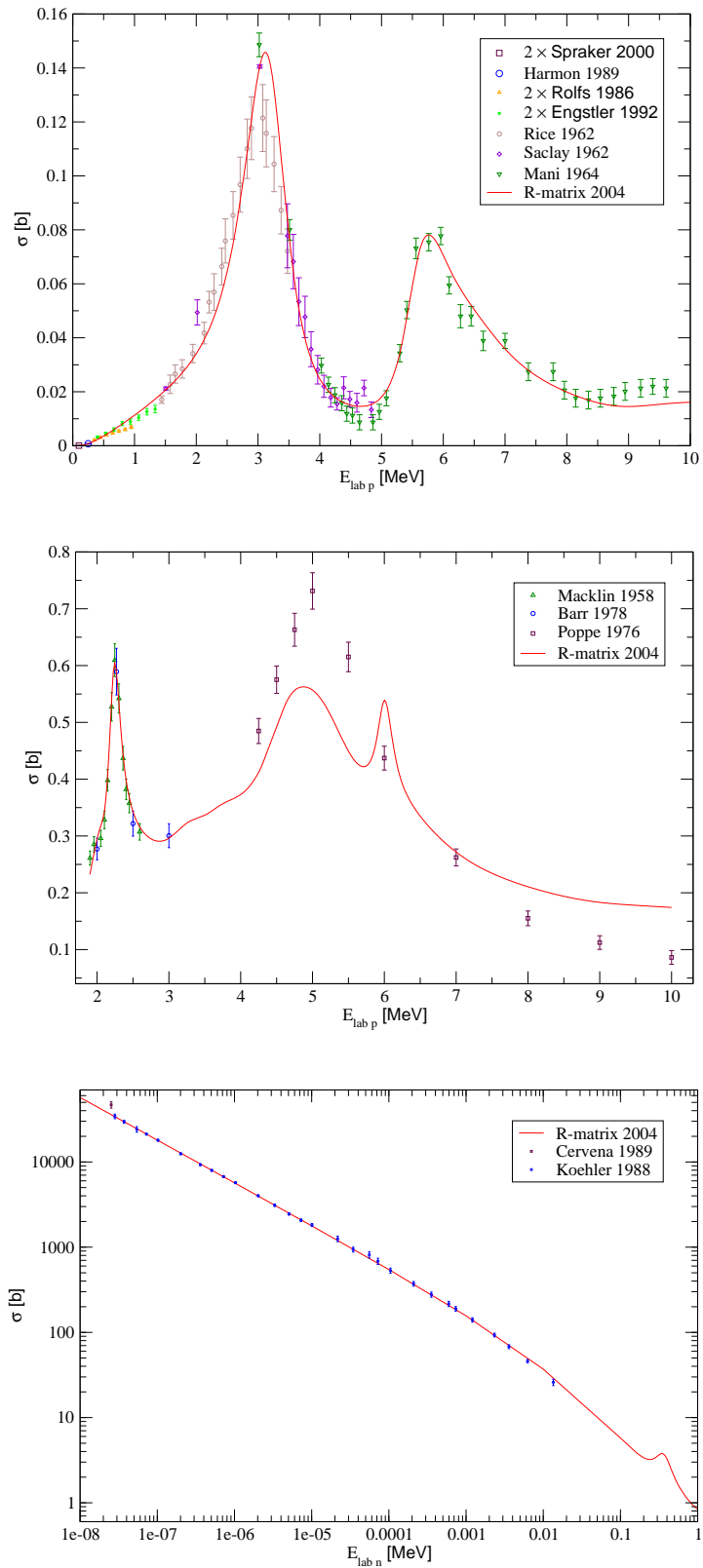}}}
 \caption{From top to bottom: Evaluated cross-sections for (a)
$^7Li (p,\alpha)$, (b) $^7Li (p,n)$ and (c) $^7Be (n,p)$.}
\end{figure*}

\begin{figure*}[p]
 \rotatebox{270}{\scalebox{0.38}{\includegraphics*[19mm,10mm][200mm,250mm]{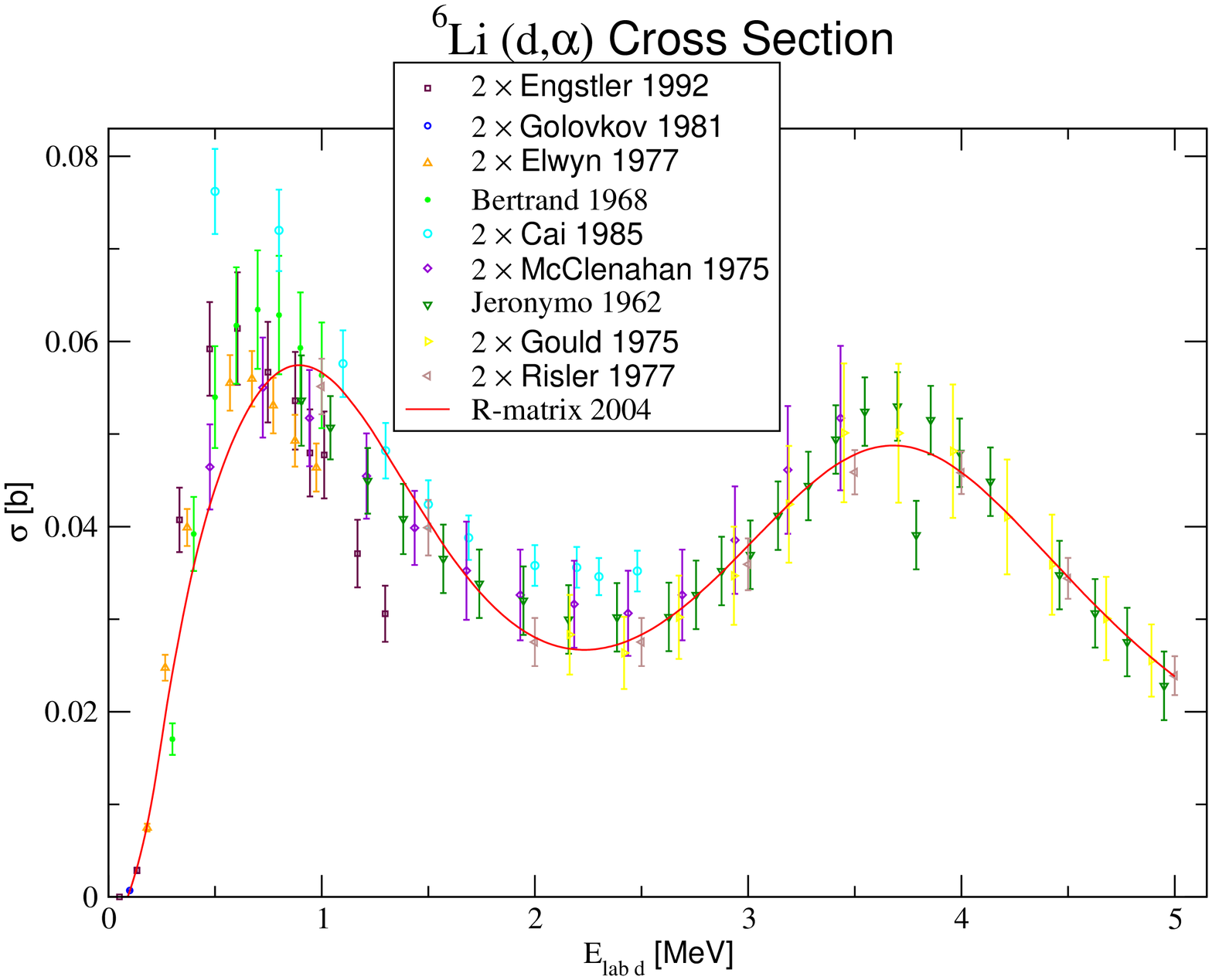}}}
\end{figure*}
\begin{figure*}[p]
 \rotatebox{270}{\scalebox{0.38}{\includegraphics*[30mm,10mm][200mm,250mm]{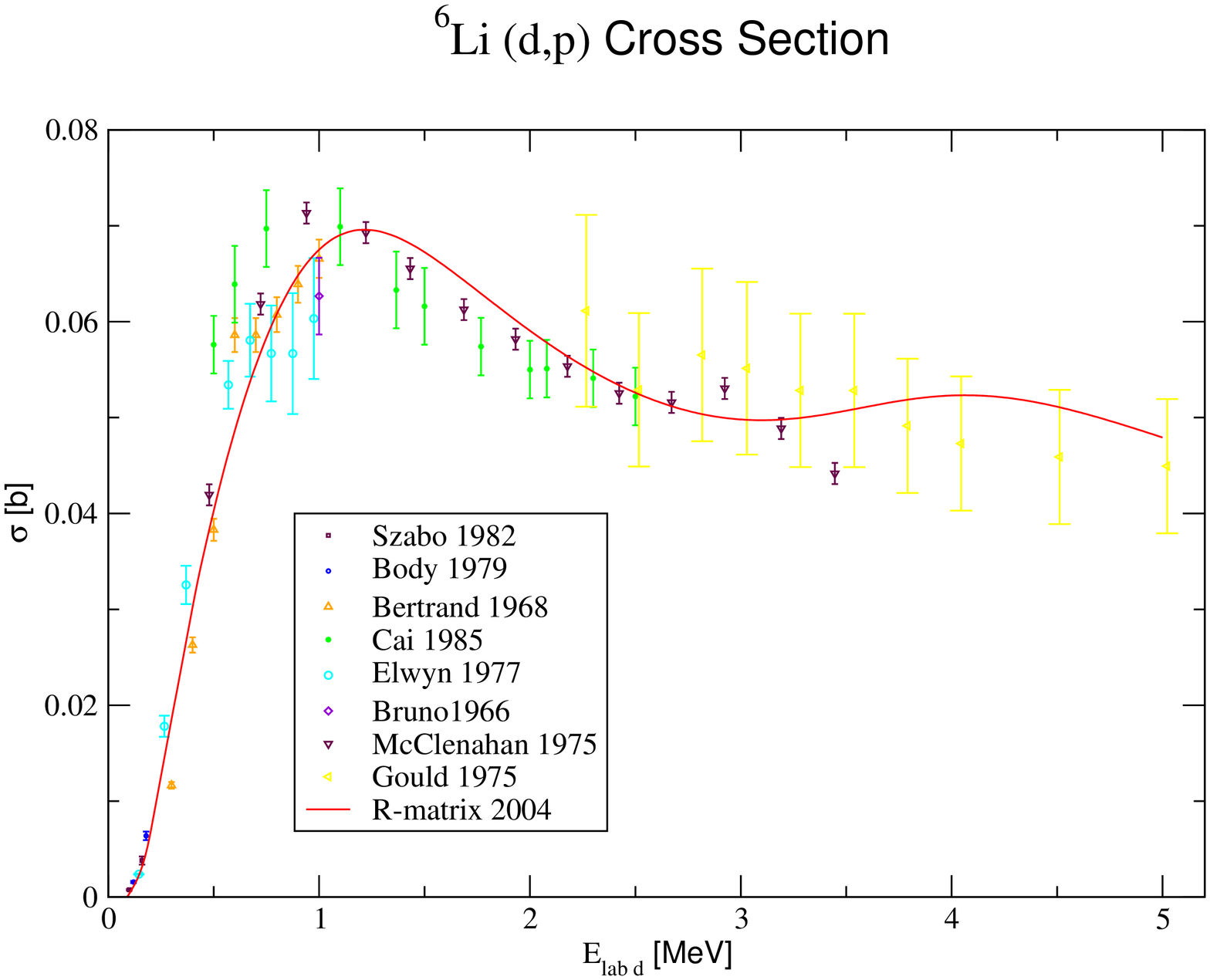}}}
\end{figure*}
\begin{figure*}[p]
 \rotatebox{270}{\scalebox{0.38}{\includegraphics*[30mm,10mm][200mm,250mm]{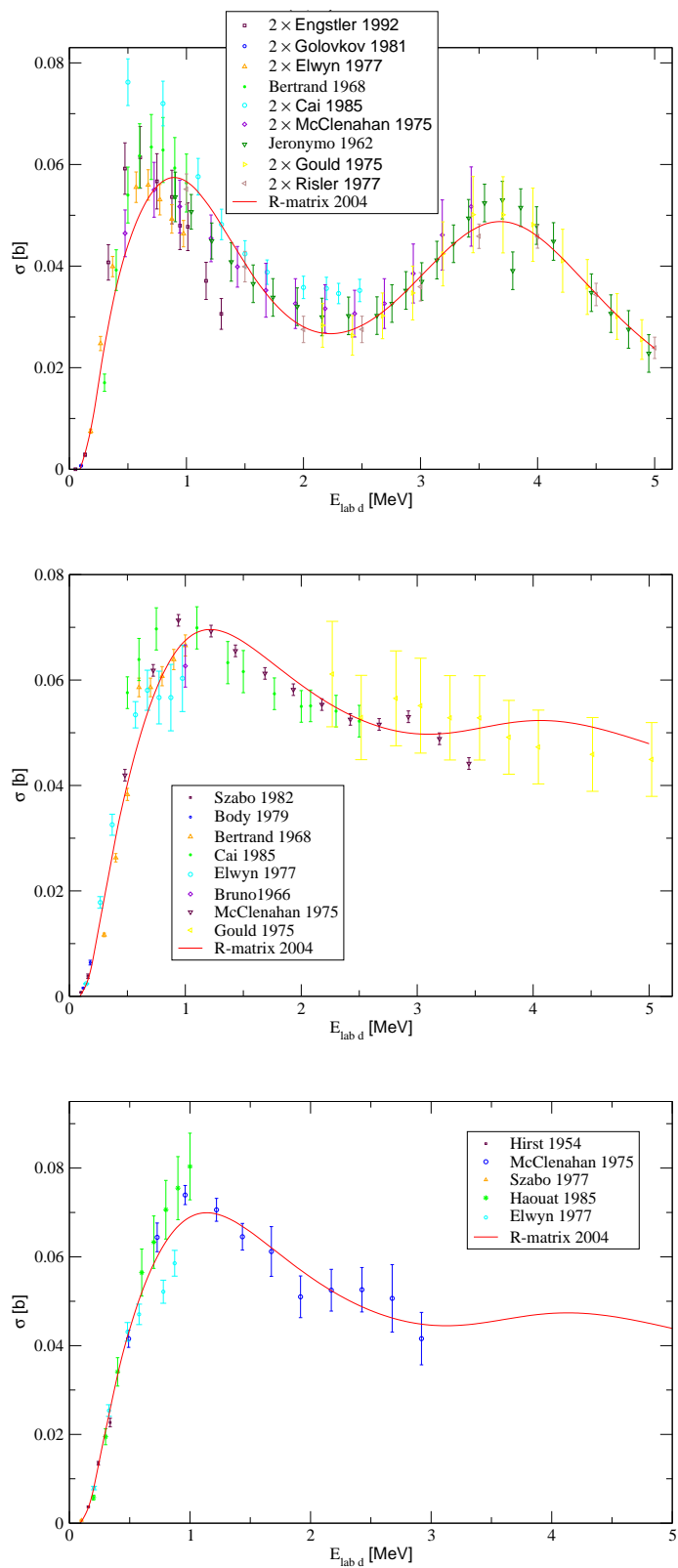}}}
 \caption{From top to bottom: Evaluated cross-sections for (a) $^6Li
(d,\alpha)$, (b) $^6Li (d,p)$ and (c) $^6Li (d,n)$.}
\end{figure*}




\bibliographystyle{aipproc}   

\bibliography{sample}

\IfFileExists{\jobname.bbl}{}
 {\typeout{}
  \typeout{******************************************}
  \typeout{** Please run "bibtex \jobname" to optain}
  \typeout{** the bibliography and then re-run LaTeX}
  \typeout{** twice to fix the references!}
  \typeout{******************************************}
  \typeout{}
 }

\end{document}